\documentclass{article}

\usepackage[preprint]{nips_2018}

\usepackage[algo2e]{algorithm2e}
\usepackage[utf8]{inputenc} 
\usepackage[T1]{fontenc}    
\usepackage{hyperref}       
\usepackage{url}            
\usepackage{booktabs}       
\usepackage{amsfonts}       
\usepackage{nicefrac}       
\usepackage{microtype}      
\usepackage{amsmath}
\usepackage{graphicx}
\usepackage{subfigure}
\usepackage{array}
\newtheorem{theorem}{Theorem}
\newtheorem{lemma}{Lemma}
\newtheorem{definition}{Definition}
\usepackage{algorithm}
\usepackage{wrapfig}
\usepackage[compact]{titlesec}
\title{Extreme Classification in Log Memory}

\author{
  Qixuan Huang\thanks{equal contribution} \\
  Department of Computer Science\\
  Rice University\\
  Houston, TX 77005 \\
  \texttt{qh5@rice.edu} \\
   \And
   Yiqiu Wang$^*$ \\
   Department of Computer Science \\
   Rice University\\
   Houston, TX 77005 \\
   \texttt{yiqiu.wang@rice.edu} \\
   \AND
  Tharun Medini\\
  Electrical and Computer Engineering\\
  Rice University\\
  Houston, TX 77005 \\
  \texttt{tharun.medini@rice.edu} \\
   \And
  Anshumali Shrivastava\\
  Department of Computer Science\\
  Rice University\\
  Houston, TX 77005 \\
  \texttt{anshumali@rice.edu} \\
}

\begin{document}
\maketitle
\begin{abstract}
  We present Merged-Averaged Classifiers via Hashing (MACH) for $K$-classification with ultra-large values of $K$. Compared to traditional one-vs-all classifiers that require $O(Kd)$ memory and inference cost, MACH only needs $O(d\log{K})$ memory while only requiring $O(K\log{K} + d\log{K})$ operation for inference ($d$ is dimensionality). MACH is a generic $K$-classification algorithm, with provably theoretical guarantees, which requires $O(\log{K})$ memory without any assumption on the relationship between classes. MACH uses universal hashing to reduce classification with a large number of classes to few independent classification tasks with small (constant) number of classes. We provide theoretical quantification of discriminability-memory tradeoff. With MACH we can train ODP dataset with 100,000 classes and 400,000 features on a single Titan X GPU, with the classification accuracy of 19.28\%, which is the best-reported accuracy on this dataset. Before this work, the best performing baseline is a one-vs-all classifier that requires 40 billion parameters (160 GB model size) and achieves 9\% accuracy.  In contrast, MACH can achieve 9\% accuracy with 480x reduction in the model size (of mere 0.3GB). With MACH, we also demonstrate complete training of fine-grained imagenet dataset (compressed size 104GB), with 21,000 classes, on a single GPU. To the best of our knowledge, this is the first work to demonstrate complete training of these extreme-class datasets on a single Titan X.
\end{abstract}

\section{Motivation}
The area of extreme multi-class classification has gained significant interest recently~\citep{choromanska2015logarithmic,prabhu2014fastxml,babbar2017dismec}. Extreme multi-class refers to the vanilla multi-class classification problem where the number of classes $K$ is significantly large. A large number of classes $K$ brings a new set of computational and memory challenges in training and deploying classifiers.

The growth in the volume and dimensionality of data is a well-known phenomenon. Concurrently, there is also significant growth in the number of classes or labels of interest. In the last decade, it has been shown that many hard AI problems can be naturally modeled as a massive multi-class problem. For example, popular NLP models predict the best word, given the full context observed so far. Such models are becoming the state-of-the-art in machine translation~\citep{sutskever2014sequence}, word embeddings~\citep{mikolov2013distributed}, question answering, etc. For large-dataset, the vocabulary size can quickly run into billions~\citep{mikolov2013distributed}.

\textbf{The Hardness Associated with the Large Number of Classes}:
Due to the popularity of word classification, it is not difficult to find public datasets with a large number of classes. Microsoft released ODP data~\citep{choromanska2015logarithmic}, where the task is to predict the category for each document (see Section~\ref{sec:data}), has around 100,000 classes. The favorite fine-grained Imagenet dataset has over 21,000 classes.

{\bf Deployment Cost:} The computational, as well as the memory requirements of the classifiers, scale linearly with the number of classes $K$. For $d$ dimensional dataset the memory required by simple linear models, such as logistic regression, is $O(Kd)$, which is a significant cost for deployment. For example, for ODP dataset, with 400,000 features and 100,000 classes, the model size of simple logistic regression is $4 \times 10^{10} \times 8$ bytes, assuming 32 bits double, or $160$ gigabytes just to store the model (or parameters). Such a large model size will run out of any GPU memory, and even deploying the model is out of the scope of the main memory of any ordinary machine. The situation naturally applies to deep networks with softmax output layer where the final layer is a known memory bottleneck. Furthermore, inference with linear classifier requires $K$ inner products of size $d$ each, totaling $O(dK)$ multiplications just to predict one instance. Such high inference cost is near-infeasible in many online and latency-critical applications.

{\bf Training Cost:} Training with large parameter space, such as 40 billion in case of ODP dataset, is always an issue. The iterations are not parallelizable.  With a large number of parameters and sequential updates, the convergence is time-consuming.

As argued, due to significant model sizes (160GB or higher) we cannot train ODP dataset on a single GPU.   All existing methods use either distributed cluster or large server machines with a significant amount of main memory (RAM). Even with a reasonably large machine, we are unable to run popular methodologies for extreme classification. See our experimental sections for details.
\vspace{-0.1in}
\subsection{Memory-Computations Tradeoff: Prior Art}
\vspace{-0.1in}

In the context of multi-class classification, reducing the prediction time at the cost of increased model size was studied recently. There are two main lines of work in this regard. It was first observed in ~\citep{shrivastava2014asymmetric} that the prediction cost, of multi-class classification problem, can be reduced to a maximum inner product search instance which admits faster solution using locality sensitive hashing. ~\citep{vijayanarasimhan2014deep} demonstrated this empirically in the context of deep networks. However, this reduction comes at the cost of increased memory required for storing hash tables, along with $O(Kd)$ for the parameters. ~\citep{daume2016logarithmic}, instead of hash tables, uses smartly learned trees to prune down the search space quickly. However,  they still require storing all the parameters and the use of trees doubles the overall space complexity. Furthermore, the parallelism is hurt in the inference time due to the tree traversals.


Memory is the primary barrier to extreme classification both during training and testing. Unfortunately, reducing memory in extreme scale classification has not received enough formal treatment. There are, however, several tricks to reduce memory in multi-class classification. The high-level idea is to compress the parameter space, and whenever the parameters are needed, they are reconstructed back. Notable approaches include hashed nets~\citep{chen2015compressing}, Huffman compression~\citep{han2015deep}, etc. But compression and decompression for accessing any parameter lead to costly overheads. Furthermore, most of the compression methods only lead to around a constant factor improvement without any formal information theoretic guarantees.

{\bf A Note on Dimensionality Reduction:}  Dimensionality reduction techniques such as random projections and feature hashing~\citep{weinberger2009feature} are orthogonal ideas to reduce computations and memory by decreasing the factor $d$. On the contrary, this paper is focused on reducing the factor $K$. All the claims  of the paper are valid assuming $d$ is the optimum dimensionality which is sufficient for learning.

{\bf A Note on Sparsity:} One effective way to reduce model size is to enforce sparsity in the parameters. Sparsity does reduce the effective model size during deployment. However, to ensure that the training algorithm never exceeds the memory limit, we have to resort to some thresholding which slows down the convergence, often prohibitive for training massive datasets. Also, sparsity is an additional assumption about the structure of the problem and our proposal is free of any such assumption.


{\bf Difference from Multi-Label and Structured Predictions:} There are other tasks with a large number of classes such as multi-label learning~\citep{prabhu2014fastxml,yu2014large} and structured prediction~\citep{bakir2007predicting}. If we allow multiple outputs for each example, then every combination can easily give us a potentially large number of classes. However, the classes are not independent, and the structure can be exploited to reduce both computations and memory~\citep{jasinska2016log}.

{\bf Our Focus:} We do not make any assumptions on the classes. Our results are for any generic $K$-class classification without any relations, whatsoever, between the classes. For the sake of readability,  we use standard logistic regression as a running example for our technique and experimentations.  However, the results naturally extend to any $K$-class classifier, including deep networks. Our idea takes the existing connections between extreme classification (sparse output) and compressed sensing, pointed out in~\cite{hsu2009multi}, to an another level in order to avoid storing costly sensing matrix. We comment more on it in section~\ref{sec:CSHH}

Most existing methods, reduce inference time at the cost of increased memory and vice versa.  It is long known that there is a tradeoff between computations, memory, and accuracy.  However, in the context of multi-class classification, even with approximations, there is a need for methodologies that reduce both computation and memory simultaneously, compared to $O(Kd)$, while providing theoretical guarantees of the tradeoffs. This work presents such result for $K$-class classification.

\subsection{Our Contributions:}
We propose a simple hashing based divide and conquer algorithm MACH (Merged-Average Classification via Hashing) for $K$-class classification, which only requires $O(d\log{K})$ model size (memory) instead of $O(Kd)$ required by traditional linear classifiers. MACH also provides computational savings by requiring only $O(d\log{K} + K\log{K})$ multiplications instead of $O(Kd)$. Furthermore, the training process of MACH is embarrassingly parallelizable.

Overall, MACH uses a 2-universal random hash function to assign classes to a small number of buckets. The large output classification problem is reduced to a small output space problem. A small classifier is built on these merged buckets (or meta-classes). Using only logarithmic (in $K$) such independent classifiers, MACH can discriminate any pair of classes with high probability.

We provide strong theoretical guarantees quantifying the tradeoffs between computations accuracy and memory. In particular, we show that in $\log{\frac{K}{\sqrt{\delta}}}d$ memory MACH can discriminate between any two pair of classes with probability $1 - \delta$. Our analysis provides a novel treatment for approximation in classification via a property call distinguishability between any pair of classes. Our novel formalism of approximate multi-class classification and its strong connections with compressed sensing and heavy hitters problem could be of independent interest in itself.

MACH achieves 19.28\% accuracy on the ODP dataset which is the best-reported accuracy seen so far on this dataset. The best previous accuracy on this partitions was only 9\%~\citep{daume2016logarithmic}. Furthermore,  MACH only requires a single GPU (Nvidia Pascal Titan X). The model size with MACH is mere around 1.2GB, to achieve around 15\% accuracy compared to around 160GB with the one-vs-all classifier that gets 9\% and requires high-memory servers to process heavy models.


Sequential training of MACH on a single Titan X GPU takes only 7 hours on the ODP dataset. As MACH is trivially parallelizable,  the same training time becomes 17 minutes with 25 GPUs. Similarly, we can train linear classifiers on fine-grained imagenet features, with 21000 categories,  in around 23 hours on a single titan X and around an hour using 20 GPUs.

To the best of our knowledge, this is the first work to demonstrate training of linear classifiers over 100,000 classes on a single GPU, and the first demonstration of training fine-grained Imagenet, with 21000 classes,  on a single GPU.  These new results illustrate the power of MACH for extreme-classification scenarios.
\section{Background}
We will use the standard [] for integer range, i.e., $[l]$ denotes the integer set from 1 to $l$: $[l]= \{1, \ 2,\cdots , \ l \}$. We will use the standard logistic regressions settings. The data $D$ is given by $D = (x_i, \ y_i)_{i=1}^{N}$. $x_i \in \mathbb{R}^d$ will be $d$ dimensional features and $y_i \in \{1,\ 2,\cdots ,\ K \}$, where $K$ denote the number of categories in the multi-class classification problem. We will drop the subscript $i$ for simplicity whenever we are talking about a generic data and only use $(x, \ y)$. Given an $x$, we will denote the probability of $y$ (label) taking the value $i$, under the given classifier model, as $p_i = Pr(y = i|x)$.

\vspace{-0.1in}
\subsection{2-Universal Hashing}
\label{sec:2univ}
\vspace{-0.1in}

{\bf Definition:} A randomized function $h: [l] \rightarrow [B]$ is 2-universal if for all, $i, j \in [l]$ with $i \ne j$, we have the following property for any $z_1, z_2 \in [k]$
\begin{align}
  Pr(h(i) = z_1 \mbox{ and } h(j) = z_2 ) = \frac{1}{B}
\end{align}
~\citep{carter1977universal} showed that the simplest way to create a 2-universal hashing scheme is to pick a prime number $p \ge B$, sample two random numbers $a, \ b$ uniformly in the range $[0,p]$ and compute $h(x) = ((ax+b)~{\bmod  ~}p)~{\bmod  ~}B$.

The fastest way to get 2-universal hashing with b-bit range is to sample a random odd number in the range $[2^{32}]$ and simply perform $h(x) = (unsigned) ~ ax ~ \& ~ (2^b -1)$. This is because in $ax$ is already $ax ~ \bmod ~ 2^{32}$ due to integer overflow. Taking another $\bmod ~ 2^b$ can be done using bitwise \& operation.

\vspace{-0.1in}
\section{Merged-Averaged Classifiers via Hashing (MACH)}
\vspace{-0.05in}
\label{sec:proposal}
\vspace{-1mm}
Formally, we use $R$, independently chosen, 2-universal hash functions $h_i: [K] \rightarrow [B]$, $i = \{1,\ 2,\cdots,\ R \}$. Each $h_i$ uniformly maps the classes (total $K$ classes) into one of the $B$ buckets. $B$ and $R$ are our parameters that we can tune to trade accuracy with both computations and memory.  $B$ is usually a small constant like $10$ or $50$. Given the data $\{x_i,\ y_i\}_{i=1}^{N}$, it is convenient to visualize, that each hash function $h_j$, transforms the data $D$ into $D_j = \{x_i,\  h_j(y_i)\}_{i=1}^{N}$. We do not have to materialize the hashed class values for all small datasets, we can simply access the class values through $h_j$. We then train $R$ classifiers (models) on each of these $D_j$'s to get $R$ models $M_j$s. This concludes our training process. Note each $h_j$ is independent. Training $R$ classifiers is trivially parallelizable across $R$ machines or GPUs.

We need a few more notations. Each meta-classifier can only classify among the merged meta-classes. Let us denote the probability of the meta-class $b \in [B]$, with the $j^{th}$ classifier with capitalized $P^j_{b}$. If the meta-class contains the class $i$, i.e. $h_j(i) = b$, then we can also write it as $P^j_{h_j(i)}$.  We are now ready to describe the prediction phase.

During prediction, we show (in Section~\ref{sec:analysis}) that the following expression is a good estimator of the probability of class $i$, given any feature vector $x$.
\begin{equation}\label{eq:probabilityclassi}
  Pr(y = i| x) \ \ = \ \ \frac{B}{B-1}\bigg[\frac{1}{R}\sum_{j=1}^R P^j_{h_j(i)}(x) - \frac{1}{B}\bigg],
\end{equation}
here $P^j_{h_j(i)}(x)$ is the predicted probability of meta-class $h_j(i)$ under the $j^{th}$ model ($M_j$), for  given $x$.
Thus, our classification rule is given by $\arg \max_i Pr(y = i|x) = \arg \max_i \sum_{j=1}^R P^j_{h_j(i)}(x).$  In other words, during prediction, for every class $i$, we get the probability assigned to associated meta-class $P^j_{h_j(i)}(x)$, by the $j^{th}$ classifier on $x$. We compute the sum of these assigned probabilities, where the summation is over $j$. Our predicted class is the class with the highest value of this sum.  The overall details are summarized in algorithm~\ref{alg:train} and~\ref{alg:predict}.




\begin{minipage}[b]{6cm}
  \vspace{0pt}  
  \begin{algorithm}[H]
  \SetAlgoNoLine
  \KwData{$D = (X, Y) = (x_i, \ y_i)_{i=1}^{N}$. $x_i \in \mathbb{R}^d$ $y_i \in \{1,\ 2,\cdots ,\ K \}$}
  \SetKwInOut{Input}{Input}\SetKwInOut{Output}{Output}
  \Input{$B, R$}
  \Output{$R$ trained multi-class classifiers}
  initialize $R$ 2-universal hash functions $h_1, h_2,...h_R$\\
  initialize $result$ as an empty list\\
  \For{$i\leftarrow 1$ \KwTo $R$}{
  	$Y_{h_i} \leftarrow h_i(Y)$\\
    $M_i = trainLogistic(X, Y_{h_i})$\\
    Append $M_i$ to $result$
  }
  \Return{result}

  \caption{Train}
  \label{alg:train}
\end{algorithm}
\hspace{-3.2in}
\end{minipage}%
\hspace{1cm}
\begin{minipage}[b]{6cm}
  \vspace{0pt}
  \begin{algorithm}[H]
  \SetAlgoNoLine
  \KwData{$D = (X, Y)= (x_i, \ y_i)_{i=1}^{N}$. $x_i \in \mathbb{R}^d$ $y_i \in \{1,\ 2,\cdots ,\ K \}$}
  \SetKwInOut{Input}{Input}\SetKwInOut{Output}{Output}
  \Input{$M = {M_1, M_2,...,M_R}$}
  \Output{$N$ predicted labels}
  \BlankLine
  load $R$ 2-universal hash functions $h_1, h2,...h_R$ used in training\\
  initialize $P$ as a an empty list\\
  initialize $G$ as a $(|N| * K)$ matrix\\
  \For{$i\leftarrow 1$ \KwTo $R$}{
  	$P_i = getProbability(X, M_i)$\\
    Append $P_i$ to $P$
  }
  \For{$j\leftarrow 1$ \KwTo $K$}{
  \tcc{$G[:, j]$ indicates the $j$th column in matrix $G$}
   	$G[:, j] = (\sum_{r=1}^{R}{P_r[:, h_r(j)]}) / R$
  }
  \Return{argmax(G, axis=1)}

  \caption{Predict}
  \label{alg:predict}
\end{algorithm}

\end{minipage}

Clearly, the total model size of MACH is $BRd$ to store $R$ models of size $Bd$ each.  The prediction cost requires $RBd$ multiplications to get meta probabilities, followed by $KR$ to compute equation~\ref{eq:probabilityclassi} for each of the classes, the maximum can be calculated on the fly. Thus, the total cost of prediction is $RBd + KR$.

\subsection{Theoretical Analysis}
\label{sec:analysis}


Classification algorithms such as logistic regression and deep networks models the probability $Pr(y = i|x) = p_i$. For example, the famous softmax or logistic modelling uses $Pr(y = i|x) = \frac{e^{\theta_i\cdot x}}{Z}$, where $Z$ is the partition function. With MACH, we use $R$ 2-universal hash functions. For every hash function $j$, we instead model $Pr(y = b|x) = P^j_b$, where $b \in [B]$. Since $b$ is a meta-class, we can also write $P^j_b$ as
\begin{align}\label{eq:metaprob}
  P_b^j = \sum_{i:h_j(i) = b} p_i; \ \ \ \ \ \  \ 1= \sum_{i=1}^K p_i = \sum_{b \in [B]} P_b^j \ \ \ \ \  \forall j
\end{align}

With the above equation, given the $R$ classifier models, an unbiased estimator of $p_i$ is:
\begin{theorem}
  \begin{align}\label{eq:theoremunbiased}
    \mathbb{E}\bigg[\frac{B}{B-1}\bigg[\frac{1}{R}\sum_{j=1}^R P^j_{h_j(i)}(x) - \frac{1}{B}\bigg]\bigg] &= Pr\bigg(y = i \bigg| x \bigg) = p_i
  \end{align}
\end{theorem}
{\bf Proof:} Since the hash function is universal, we can always write $$P^j_{h(i)} = p_i + \sum_{k \ne i} {\bf 1}_{h(k) = h(i)}p_k,$$ where ${\bf 1}_{h(k) = h(i)}$ is an indicator random variable with expected value of $\frac{1}{B}$.  Thus $E(P^j_{h(i)}) = p_i + \frac{1}{B}\sum_{k \ne  i} p_k  = p_i + (1 - p_i) \frac{1}{B}$. This is because the expression $\sum_{k \ne  i} p_k = 1 - p_i$ as the total probability sum up to one (assuming we are using logistic type classfiers).  Simplifying, we get $p_i = \frac{B}{B-1}(E(P^j_{h(i)}(x) - \frac{1}{B}) $. It is not difficult to see that this value is also equal to $\mathbb{E}\bigg[\frac{B}{B-1}\bigg[\frac{1}{R}\sum_{j=1}^R P^j_{h_j(i)}(x) - \frac{1}{B}\bigg]\bigg]$ using linearity of expectation and the fact that $E(P^j_{h(i)})   = E(P^k_{h(i)}) $ for any $j \ne k$.

For a $d$ dimensional dataset (or $d$ non-zeros for sparse data), the memory required by a vanilla logistic regression model (or any linear classifier) is $O(Kd)$. $O(Kd)$ is also the computational complexity of prediction. With MACH, the computational, as well as the memory complexity, is dependent on $B$ and $R$ roughly as $O(BR + KR)$. To obtain significant savings, we want $BR$ to be significantly smaller than $Kd$. We next show that $BR \approx O(\log{K})$ is sufficient for identifying the final class with high probability.

\begin{definition}
{\bf Indistinguishable Class Pairs:} Given any two classes $c_1$ and $c_2$  $\in [K]$, they are indistinguishable under MACH if they fall in the same meta-class for all the $R$ hash functions, i.e., $h_j(c_1) = h_j(c_2)$ for all $j \in [R]$.
\end{definition}

Otherwise, there is at least one classifier which provides discriminating information between them. Given that the only sources of randomness are the independent 2-universal hash functions, we can have the following lemma
\begin{lemma}
 MACH with $R$ independent $B$-class classifier models, any two original classes $c_1$ and $c_2$  $\in [K]$ will be indistinguishable with probability at most
\begin{align}\label{eq:upperboundonepair}
  Pr(\text{classes $i$ and $j$  are indistinguishable}) \le \bigg(\frac{1}{B}\bigg)^{R}
\end{align}
\end{lemma}

There are total $\frac{K(K-1)}{2} \le K^2$ possible pairs, and therefore, the probability that there exist at least one pair of classes, which is indistinguishable under MACH is given by the union bound as
\begin{align}\label{eq:upperboundallpair}
  Pr(\exists \ \ \text{an indistinguishable pair}) \le  K^2 \bigg(\frac{1}{B}\bigg)^{R}
\end{align}

Thus, all we need is $K^2 \bigg(\frac{1}{B}\bigg)^{R} \le \delta$ to ensure that there is no indistinguishable pair with probability $\ge 1 - \delta$. Overall, we get the following theorem:
\begin{theorem}\label{}
  For any $B$, $R = \frac{2\log{\frac{K}{\sqrt{\delta}}}}{\log{B}}$, guarantees that all pairs of classes $c_i$ and $c_j$ are distinguishable (not indistinguishable) from each other with probability greater than $1 - \delta$.
\end{theorem}

From section~\ref{sec:proposal}, our memory cost is $BRd$ to guarantee all pair distinguishably with probability $1 - \delta$, which is equal to $\frac{2\log{\frac{K}{\sqrt{\delta}}}}{\log{B}}Bd$. This holds for any constant value of $B \ge 2$. Thus, we bring the dependency on memory from $O(Kd)$ to $O(\log{K}d)$ in general with approximations. Our inference cost is $\frac{2\log{\frac{K}{\sqrt{\delta}}}}{\log{B}}Bd + \frac{2\log{\frac{K}{\sqrt{\delta}}}}{\log{B}}K$ which is $O(K\log{K} + d\log{K})$, which for high dimensional dataset can be significantly smaller than $Kd$.
\subsection{Connections with Compressed Sensing and Heavy Hitters}
\label{sec:CSHH}
Given a data instance $x$, a generic classifier  outputs the probabilities $p_i$, $i \in \{1, \ 2,..., \ K\}$. We can then use these $K$ numbers (probabilities) for inference and training. We want to essentially compress the information of these $K$ numbers to $\log{K}$, i.e., we can only keep track of $\log{K} = BR$ numbers (or measurements). Ideally, without any assumption, we cannot compress the information in $K$ numbers to anything less than $\Omega(K)$, if we want to retain all information. However, in classification, the most informative quantity is the identity of $\arg \max p_i$.

If we assume that $\max p_i \ge \frac{1}{m}$, for sufficiently small $m \le K$, which should be true for any good classifier. Then, identifying $\arg \max p_i$ with $\sum p_i  =1$ and $\max p_i \ge \frac{1}{m} \sum p_i$ is a classical approximate heavy hitter problem. Finding heavy hitters in the data stream is more than 30 years old problem with a rich literature. Count-Min sketch~\citep{cormode2005improved} is the most popular algorithm for solving heavy hitters over positive streams. Our method of using $R$ universal hashes with $B$ range is precisely the count-min sketch measurements, which we know preserve sufficient information to identify heavy hitters (or sparsity) under good signal-to-noise ratio. Thus, if $\max p_i$ (signal) is larger than $p_j$, $j \ne i$ (noise), then we should be able to identify the heavy co-ordinates (sparsity structure) in $sparsity \times \log{K}$ measurements (or memory)~\citep{candes2006near}.

The connection between compressed sensing and extreme classification was identified in prior works~\cite{hsu2009multi,dietterich1995solving}.  In~\cite{hsu2009multi}, the idea was to use a compressed sensing matrix to compress the $K$ dimensional binary indicator vector of the class $y_i$  to a real number and solve a regression problem. Although the number of required measurements is logarithmic in $K$, the memory requirement is still super linear in $K$, as we need to store and process the compressing matrix. This matrix is required for both compression as well as inference, which is prohibitive for our datasets. Furthermore, the inference cost requires solving a sparse recovery subroutine which is iterative and can be costly in itself.

We instead compress probability vectors $p_i$s. Since probabilities are additive for merged classes (OR rule of probability), we can afford a much simpler compression and estimation which is similar to count-min sketch in disguise. Thus, we only need 2-universal hashing, which requires constant memory per hash function, for guaranteeing distinguishably (discriminative classifier) between any pair of classes with high probability.

Once this is essentially count-min sketch in disguise, it opens up two other estimators min and median apart from Equation~\ref{} for $p_i$ which are discussed in the supplementary material. Our work provides the first bridge that connects two-decade-old sketching algorithms from the data streams literature with classical extreme $K$-class classification.

\vspace{-0.1in}
\section{Evaluations}
\vspace{-0.1in}
\subsection{Datasets}
\label{sec:data}
\vspace{-0.1in}

We use the two large public benchmarks from~\citep{daume2016logarithmic}: 1) ODP dataset and 2) Fine-grained Imagenet.

ODP is a multi-class dataset extracted from Open Directory Project, the largest, most comprehensive human-edited directory of the Web. Each sample in the dataset is a document, and the feature representation is bag-of-words. The class label is the category associated with the document. The dataset is obtained from~\citep{choromanska2015logarithmic}. The number of categories in this dataset is 105,033.

Imagenet is a dataset consist of features extracted from an intermediate layer of a convolutional neural network trained on the ILVSRC2012 challenge dataset. This dataset was originally developed to study transfer learning in visual tasks~\citep{oquab2014learning}. Please see~\citep{choromanska2015logarithmic} for more details. The class label is the fine-grained object category present in the image. The number of categories in this dataset is 21,841.

The statistics of the two datasets are summarized in Table~\ref{tab:datasets}. These datasets cover different domains -- images and text. The datasets are sufficiently hard and demonstrate the challenges of having large classes. There is only one published work~\citep{daume2016logarithmic} which could successfully train a standard one-vs-all linear classifier. The training was done over a sophisticated distributed machine and the training time spanned several days.

\begin{table*}[t]
\begin{center}
\begin{tabular}{ | m{1.2cm} | m{1.2cm} | m{3cm}| m{1.4cm} | m{1.4cm} | m{3cm}| }
\hline
Name & Type & \#Train / \#Test & \#Classes & \#Features & size\\
\hline
ODP & Text & 1084404 / 493014 & 105033 & 422713 & 3GB(sparse format)\\
\hline
Imagenet & Images & 12777062 / 1419674 & 21841 & 6144 & 104GB(Compressed)\\
\hline
\end{tabular}
\end{center}
\caption{Statistics of Datasets used for Evaluations}
\label{tab:datasets}
\end{table*}

All our experiments are performed on the same server with GeForce GTX TITAN X, Intel(R) Core(TM) i7-5960X 8-core CPU @ 3.00GHz and 64GB memory.
We used Tensorflow to train each individual model $M_i$ and obtain the probability matrix $P_i$ from model $M_i$. We use OpenCL to compute the global score matrix that encodes the score for all classes $[1, K]$ in testing data and perform argmax to find the predicted class. We have our codes and scripts ready for release on Github for reproducibility.

\subsection{Accuracy Baselines}

\begin{figure*}

\centering
\subfigure{\hspace{-1.5cm} \includegraphics[width=0.5\textwidth]{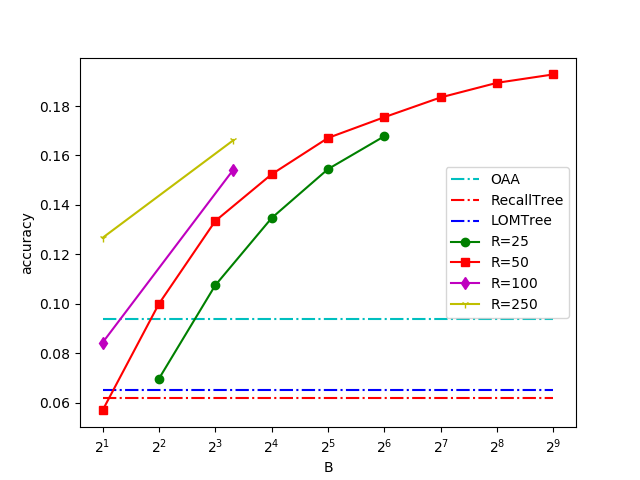}}%
\hspace{1 cm}
\subfigure{\includegraphics[width=0.5\textwidth]{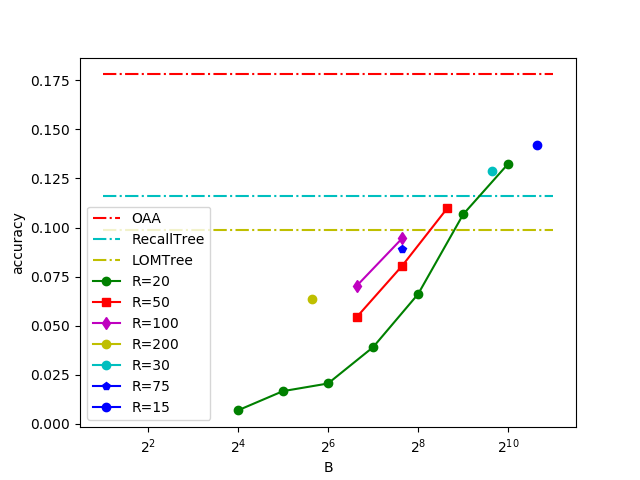}}
\caption{Accuracy Resource tradeoff with MACH (all bold lines) with varying settings of $R$ and $B$. The number of parameters are $BRd$ while the prediction time requires $KR + BRd$ operations. All the runs of MACH requires less memory than OAA. The straight line are accuracies of OAA, LOMTree and Recall Tree (dotted lines) on the same partition taken from~\citep{daume2016logarithmic}. LOMTree and Recall Tree uses more (around twice) the memory required by OAA. {\bf Left:} ODP Dataset. {\bf Right:} Imagenet Dataset}
\label{fig:accuracyplot}
\end{figure*}
On these large benchmarks, there are three published methods that have reported successful evaluations -- 1) {\bf OAA}, traditional one-vs-all classifiers, 2) {\bf LOMTree} and 3) {\bf Recall Tree}. The results of all these methods are taken from~\citep{daume2016logarithmic}.  OAA is the standard one-vs-all classifiers whereas LOMTree and Recall Tree are tree-based methods to reduce the computational cost of prediction at the cost of increased model size. Recall Tree uses twice as much model size compared to OAA. Even LOMtree has significantly more parameters than OAA. Thus, our proposal MACH is the only method that reduces the model size compared to OAA.

In addition, we experimented with three popular extreme classification algorithms with publicly available implementations: 1) {\bf DiSMEC~\cite{babbar2017dismec}}, 2) {\bf PD-sparse~\cite{yen2016pd}} and 3) {\bf FastXML~\cite{prabhu2014fastxml}}. Even on a machine with 512 GB RAM and 56 cores, DisMEC and PD-sparse couldn't run in a week and FastXML never crossed and accuracy of 1\% with various hype-parameter settings. Slow progress in DisMEC and PD-sparse is not surprising as iterative thresholding (or greedy capacity control) requires sorting of parameters based on weights, which can be prohibitive.  

We, therefore, primarily use --  1) { OAA}, traditional one-vs-all classifiers, 2) { LOMTree} and 3) { Recall Tree}-- as the baselines to contrast the accuracy-memory tradeoff provided by MACH. We directly compare against the reported accuracy as benchmarks.

\subsection{Results and Discussions}

\begin{table*}[h!]
\begin{center}
\begin{tabular}{ | p{1.2cm} | p{1.5cm} | p{2cm}| p{2cm} | p{3cm} | p{2cm} | }
\hline
Dataset & (B, R) & Model size Reduction & Training Time & Prediction Time per Query & Accuracy \\
\hline
ODP & (32, 25) & 125x & 7.2hrs & 2.85ms & 15.446\% \\
\hline
Imagenet & (512, 20) & 2x & 23hrs & 8.5ms & 10.675\% \\
\hline
\end{tabular}
\end{center}
\vspace{-0.1in}
\caption{Wall Clock Execution Times and accuracies for two runs of MACH on a single Titan X.}
\vspace{-0.1in}
\label{tab:executionTime}
\end{table*}

We run MACH on these two datasets and plot accuracy as a function of varying $B$ and $R$ in Figure~\ref{fig:accuracyplot}. $B$ and $R$ are two knobs in MACH to balance resource and accuracy. We used plain logistic regression classifier, i.e., cross entropy loss without any regularization, in the Tensorflow environment~\citep{abadi2016tensorflow}. We use the unbiased estimator given by Equation~\ref{eq:probabilityclassi} for inference. Comparison to min and median estimators in shown in supplementary material.


The plots show that for ODP dataset MACH can even surpass OAA achieving 18\% accuracy while the best-known accuracy on this partition is only 9\%. LOMtree and Recall Tree can only achieve 6-6.5\% accuracy. It should be noted that with 100,000 classes, a random accuracy is $10^{-5}$. Thus, the improvements are staggering with MACH. Even with $B=32$ and $R=25$, we can obtain more than 15\% accuracy with $\frac{105,000}{32 \times 25} = 120$ times reduction in the model size. Thus, OAA needs 160GB model size, while we only need around 1.2GB. To get the same accuracy as OAA, we only need $R=50$ and $B=4$, which is a 480x reduction in model size requiring mere 0.3GB model file.

We believe that randomization and averaging in MACH cancel the noise and lead to better generalization. Another reason for the poor accuracy of other baselines could be due to the use of VW~\citep{langford2007vowpal} platforms. VW platform inherently uses feature hashing that may lose some information in features, which is critical for high dimensional datasets such as ODP.

On Imagenet dataset, MACH can achieve around 11\% which is roughly the same accuracy of LOMTree and Recall Tree while using $R=20$ and $B=512$. With $R=20$ and $B=512$, the memory requirement is $\frac{21841}{512 \times 20} = 2$ times less than that of OAA. On the contrary, Recall Tree and LOMTree use 2x more memory than OAA. OAA achieves the best result of $17\%$. With MACH, we can run at any memory budget.

\vspace{-0.1in}
\section{Conclusion}
\vspace{-0.1in}

We have shown how the problem of extreme classification with 100,000 classes and over 40 billion parameters can be run with superior accuracy on a single Titan X with mere 12GB memory limit in few hours. With more GPUs, this trivially goes down to few minutes. We exploit the power of fundamental count-min sketch algorithm from data stream literature and propose MACH. MACH requires memory that scales only logarithmic with the number of classes. Surprisingly, all we need are independent and small vanilla logistic regressions, which are massively parallelizable.  Thus, we naturally leverage all the advances in the fast implementation of logistic regression. 

Machine learning with large parameter space is becoming increasingly common. To get around the prohibitive memory and computation of large models it is common to shard these parameters across nodes and exploit the power of distributed parallelism.  However, often the overhead of distributed computing is far more than its advantages. 

In this work, we have demonstrated that distributed computing should be our last resort, and sometimes smart algorithmic solution, often randomized, can do wonders.

\newpage
\bibliography{main}
\bibliographystyle{unsrtnat}
\newpage
\section{Supplementary Material}
\subsubsection{Two more Estimators of Probabilities}
\label{sec:2est}
One we have identified that our algorithm is essentially count-min sketch in disguise, it naturally opens up two other possible estimators, in addition to Equation 2 for $p_i$ in the main paper. The popular min estimator, which is used in traditional count-min estimator given by:
\begin{equation}\label{eq:estmin}
  \hat{p_i}^{min} =  \underset{j}{\mathrm{min}} \ P^j_{h_j(i)}(x).
\end{equation} We can also use the median estimator used by count-median sketch~\cite{charikar2002finding} which is another popular estimator from the data streams literature.
\begin{equation}\label{eq:estmed}\hat{p_i}^{med} = \underset{j}{\mathrm{median}} \ P^j_{h_j(i)}(x).\end{equation}

\subsubsection{Comparison of Three Estimators}
\label{sec:estcomp}

\begin{table}[h]
\begin{center}
\begin{tabular}{ | m{1.2cm} | m{1.2cm} | m{1.2cm} | m{1.2cm}| }
\hline
Dataset & Unbiased & Min & Median \\
\hline
ODP & {\bf 15.446} & 12.212 & 14.434 \\
\hline
Imagenet & 10.675 & 9.743 & {\bf 10.713}\\
\hline
\end{tabular}
\end{center}
\caption{Classification accuracy with three different estimators from sketches (see section~\ref{sec:2est} for details). The training configuration are given in Table~\ref{tab:executionTime}}
\label{tab:estacccomp}
\end{table}

Finally, we also evaluated the effect of two different estimators, the min and the median, shown in Section~\ref{sec:2est}. We use the same trained model from Table~\ref{tab:executionTime} and use three different estimators, given by Eq~\ref{eq:probabilityclassi}, \ref{eq:estmin} and \ref{eq:estmed} respectively, for estimating the probability. The estimation is followed by argmax to infer the class label. It turns out that our unbiased estimator in Equation~\ref{eq:probabilityclassi} performs overall the best. Median is slightly better on imagenet data and poor on ODP dataset. Min estimator leads to poor results on both of them. The accuracy results are summarized in Table~\ref{tab:estacccomp}

\subsection{Running Times and Parallelism}

All our experiments used a single Titan X to run the combinations shown in Figure~\ref{fig:accuracyplot}. There has been no prior work that has reported the run of these two datasets on a single Titan X GPU with 12GB memory. This is not surprising due to memory limit.

With MACH, training is embarrassingly easy, all we need is to run $R$ small logistic regression which is trivially parallelizable over $R$ GPUs. In Table~\ref{tab:executionTime}, we have compiled the running time of some of the reasonable combination and have shown the training and prediction time. The prediction time includes the work of computing probabilities of meta-classes followed by sequential aggregation of probabilities and finding the class with the max probability.


It should be noted that if we have $R$ machines then this time would go down by factor of $R$, due to trivial parallelism of MACH because all classifiers are completely independent. Although, we cannot compare the running time with any previous methods because they use different systems. Nonetheless, the wall clock times are significantly faster than the one reported by RecallTree, which is optimized for inference.
The most exciting message is that with MACH we can train this intimidating datasets on a single GPU. We hope MACH gets adopted in practice.
\end{document}